\documentclass[aps,pre,epsfigure,twocolumn]{revtex4-1}
\usepackage[colorlinks=true,linkcolor=blue,urlcolor=blue,citecolor=blue,pdfusetitle]{hyperref}
\usepackage[utf8]{inputenc}
\usepackage[english]{babel}
\usepackage{amsmath}
\usepackage[caption = false]{subfig}
\usepackage{graphicx,epstopdf}
\usepackage{blindtext}
\usepackage{lipsum}
\usepackage{amsfonts}
\usepackage{bbm}
\usepackage{amssymb}
\usepackage{enumerate}
\usepackage{color}
\usepackage{latexsym}
\usepackage{physics}
\usepackage{times,txfonts}

\begin{document}

\title{Steady-state charging of quantum batteries via dissipative ancillas} 

\author{F. H. Kamin}
\email{f.hatami@uok.ac.ir}
\affiliation{Department of Physics, University of Kurdistan, P.O.Box 66177-15175 , Sanandaj, Iran}

\author{M. B. Arjmandi}
\email{arjmandi94@gmail.com}
\affiliation{Department of Physics, University of Isfahan, P.O. Box 81746-7344, Isfahan, Iran}

\author{S. Salimi}
\email{shsalimi@uok.ac.ir}
\affiliation{Department of Physics, University of Kurdistan, P.O.Box 66177-15175 , Sanandaj, Iran}

\begin{abstract}
We investigate the steady-state charging process of a single-cell quantum battery embedded in an N-cell star network of qubits, each interacting with a fermion reservoir, collectively and individually in equilibrium and non-equilibrium scenarios, respectively. We find an optimal steady-state charging in both scenarios, which grows monotonically with the reservoirs' chemical potential and chemical potential difference. Where the high base temperature of the reservoirs has a destructive role in all parameter regimes. We indicate that regardless of the strength of the non-equilibrium condition, the high base chemical potential of the battery's corresponding reservoir can significantly enhance the charging process. On the other hand, a weak coupling strength can strongly suppress the charging. Consequently, our results could counteract the detrimental effects of self-discharging and provide valuable guidelines for enhancing the stable charging of open quantum batteries in the absence of an external charging field.
\end{abstract}

\maketitle

\section{Introduction}
In recent years, there has been a significant increase in research on the miniaturization of technological devices, leading to the emergence of promising technologies such as quantum batteries i.e., quantum systems that supply energy \cite{Alicki:1,campaioli2023}. Quantum batteries represent a vital field of research that concerns designing optimal energy storage protocols for transferring to quantum devices \cite{Andolina:2,Farina:1}. They can be charged with higher power and at a faster rate compared to their classical counterparts by using quantum resources such as coherence and entanglement \cite{Alicki:1,Campaioli:1,Andolina:1,Kamin:2,Gyhm:1, Arjmandi:1}. Already, researchers have proposed various protocols to improve the charging process of quantum batteries \cite{Binder:1,Rossini:1,Ferraro:1,Le:1,Crescente:1,Ghosh:1}. Additionally, there has been progress towards experimental implementation \cite{James:1, Hu:1,Joshi:1,wenniger:1,Gemme:1,Zheng:1,Huang:1}.

A variety of different scenarios have been envisioned for quantum batteries as a set of two-level systems. In such settings, the highest possible energy extraction via cyclic unitary processes is a central factor. This involves the use of ergotropy and finding the most effective unitary operation to bring the system to its corresponding passive state \cite{Allahverdyan:1,pusz:1}, where the battery is considered empty. To the best of our knowledge, research on closed \cite{Ferraro:1,Crescente:1,Caravelli:1,Le:1,Zhang:1} and open quantum batteries \cite{Kamin:1,Tabesh:2,Quach:1,Zakavati:1,Santos:1,Santos:2,Arjmandi:2,Ghosh:1,Konar:1,sen:1,Kamin_2023} has made significant progress. In the first scenario, the battery and charger are not impacted by the environment, including Dicke quantum batteries \cite{Ferraro:1,Crescente:1}, Random quantum batteries \cite{Caravelli:1}, spin-chain quantum battery \cite{Le:1}, a quantum battery made of non-interacting two-level atoms that can be fully charged using a harmonic charging field \cite{Zhang:1}, and so on. Implementing second-case quantum batteries in an open quantum system framework is important to discuss and ensure feasibility. Research on open quantum batteries has mainly focused on analyzing different environmental models. A crucial aspect of this case is developing efficient strategies to mitigate environmental damage to quantum battery performance. Recent studies suggest that strong couplings between a quantum battery and its environment may cause non-Markovian effects and improve quantum battery performance \cite{Kamin:1,Tabesh:2}. As another open quantum battery protocol, the authors boost charging power and capacity by utilizing dark states, where non-interacting spins are coupled to a reservoir \cite{Quach:1}. Furthermore, among the myriad challenges of quantum batteries, one of the most critical is the issue of self-discharging caused by environmental factors \cite{Santos:1,Santos:2,Arjmandi:2}. This process leads to the gradual loss of charge due to the intrinsic traits of the system employed as the energy storage medium, and this occurs independently of whether the battery is linked to a consumption hub. However, it is possible that the interaction with the environment becomes beneficial for work storage resources under some new approaches. Where there is no requirement for an extra charger or booster to counteract environmental damage.

Recently, there has been growing interest in the study of steady states in both equilibrium and non-equilibrium environments \cite{Latune:1,PhysRevE.102.042111,Wang:1,Fischbach:1}. Research indicates that entanglement and coherence are quantum phenomena that can persist in steady states, even when the system interacts with the environment through information, matter, and energy exchange. At equilibrium, the system collectively interacts with the environment at a finite temperature and/or chemical potential. While non-equilibrium conditions are retained by a constant temperature or chemical potential difference to drive energy or matter flow through the quantum system and environments, resulting in a steady deviation from equilibrium regime. Undoubtedly, the remaining quantum properties in steady states open a new path towards the production, protection and control of ergotropy through equilibrium and non-equilibrium regimes.

In this work, aiming at using the environment-induced capability for quantum battery charging, we focus on mostly unexplored aspects of the steady-state charging process. To this goal, we consider an N-qubit star network where a single-cell quantum battery is embedded in the center of the network and couples with an arbitrary number of ancillas. Then, we numerically investigate the steady-state charging process of the single-cell quantum battery when each cell interacts with a fermion reservoir that can exchange particles with the system in both equilibrium and non-equilibrium settings. Our investigations suggest a substantial steady-state charging when the network is coupled to fermion reservoirs. Where the chemical potential of reservoirs can cause diverse and intense effects, regardless of the initial state or coupling regime.
In the light of these results, we provide protocols for the realization of optimal open quantum batteries without an external charging field. Moreover, we propose a quantum battery capable of being charged within an environment that induces self-discharging, permanently storing this charge in a steady-state rather than temporarily. This method of energy storage eliminates the need for an external charger, rendering it highly energetically efficient. Notably, the achieved steady-state is independent of the initial state, thus enabling the selection of the most unconstrained state, such as one lacking any prior excitation preparation.        

The remainder of the paper is arranged as follows. Sect.~\ref{II} introduces the our model and provides the relevant physical quantities to characterize the behavior of the quantum battery. Then we discuss the charging process of the quantum battery for both equilibrium and non-equilibrium regime in Sect.~\ref{III} by presenting numerical results. We finally conclude in Sect.~\ref{IV} with a short summary.

\begin{figure}[t] 
	\includegraphics[scale=0.6]{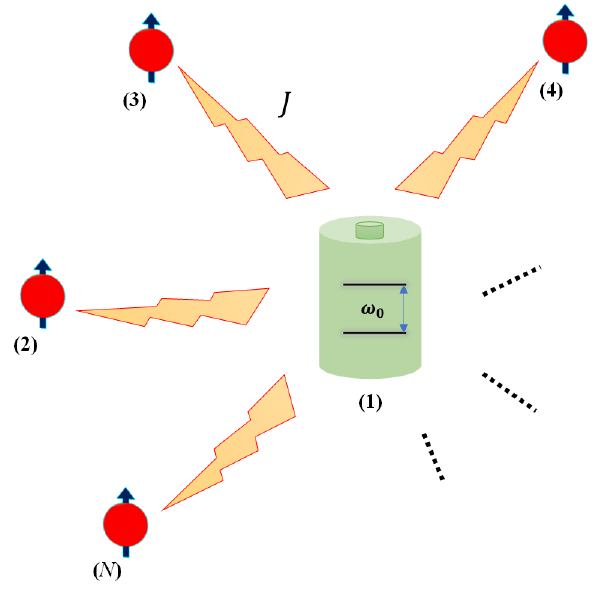}
	\caption{ A schematic diagram for the N-qubit star configuration via coupled two-level systems with transition frequencies $\omega_{0}$ connected by dipole-dipole interactions, where $J$ is the coupling strength. The qubit labeled with (1) is a central single-cell quantum battery, which interacts with all other (N-1) cells surrounding it.}
	\label{Fig.1}
\end{figure}
\section{Model and methods}\label{II}
The model under consideration is illustrated in Fig.~\ref{Fig.1}. As can be seen, each qubit (or two-level systems) is coupled by a dipole-dipole interaction to another qubit in a N-qubit star network, and quantum battery is embedded in the center of it. The system cells are assumed to be identical as a two-level system with excitation state $\vert e\rangle$ and ground state $\vert g\rangle$ and the same transition frequency $\omega_{i}=\omega_{0}~(i=1,...,N)$. The Hamiltonian of the coupled qubit system during the charging process reads ($\hbar=k_{B}=1$ in the following)
\begin{align}\label{H0array}
	H _{S} = \sum_{i=1}^{N}\omega_{0}\sigma_{z}^{i} + \sum_{i=1}^{N-1}J~(\sigma_{+}^{1}\sigma_{-}^{i+1} + \sigma_{-}^{1}\sigma_{+}^{i+1})~,
\end{align}
where $\sigma_{j}~(j=x,y,z)$ are the usual Pauli matrices and $J$ being the inter-qubit coupling strength. Here, $\sigma_{+}^{i}$ and $\sigma_{-}^{i}$ are the Pauli raising and lowering operators for $i$th qubits, respectively.

In this work, the charging process of quantum battery is non-unitary, while the maximum extractable work can be calculated by an unitary cyclic evolution when we couple the battery to some consumption hub as following form \cite{Allahverdyan:1} 
\begin{equation}\label{4}
	\mathcal{E}(\rho_{QB})=tr(\rho_{QB}H_{QB})-\min_{U}tr(U\rho_{QB}U^{\dagger} H_{QB})~,
\end{equation}
where the optimization is performed for unitary operators $U$ in the particular unitary group $\text{SU}(d)$. Therefore, the maximum amount of work that can be extracted from a system is determined by
\begin{equation}
	\mathcal{E}(\rho_{QB})=tr(\rho_{QB}H_{QB})-tr(\sigma_{QB}H_{QB})~.
\end{equation}
Such an amount of energy is well known as \textit{ergotropy} \cite{Allahverdyan:1}. Here, $\sigma_{QB}$ is the passive state (the empty battery state), since no work can be extracted from it or $tr(\sigma_{QB}H_{QB})\leq tr(U\sigma_{QB}U^{\dagger}H_{QB})$. A density matrix $\rho$ is passive if and only if it is diagonal in the basis of the Hamiltonian, and its eigenvalues are non-increasing with the energy \cite{pusz:1}.
We define $\mathcal{E}_{\text{max}}$ as the maximum achievable work of the quantum battery. Thus, the ergotropy can be normalized with respect to the maximum amount of work and denoted as $\mathcal{W}=\mathcal{E}/\mathcal{E}_{\text{max}}$. Subsequently, the term ergotropy refers to the normalized ergotropy for the sake of brevity throughout this paper. 

\section{Results}\label{III}
In this paper, we focus on situations where the ergotropy storage resources can survive in the steady-state solutions of dynamics in the presence a arbitrary number of dissipative ancillas by different scenarios. In the following, we investigate the steady-state charging process of a single-cell quantum battery first for the equilibrium setup and then move on to the non-equilibrium scenario. 

\subsection{equilibrium case}
\begin{figure}[t] 
	\includegraphics[scale=0.55]{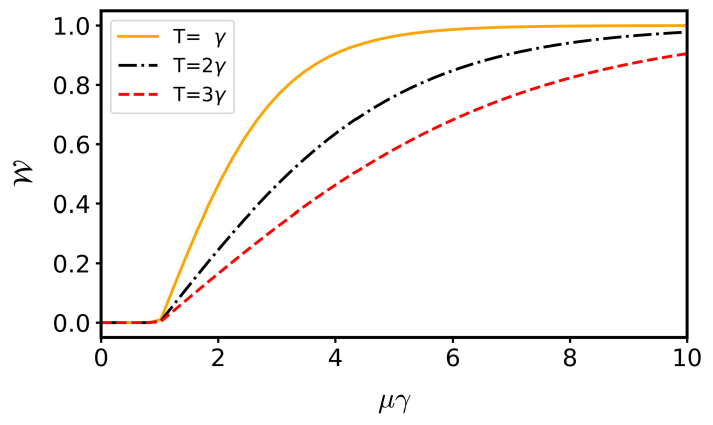}
	\caption{The ergotropy for the equilibrium case as a function of $\mu\gamma$ for different values of the temperature $T$. The parameters are set as $N=3$, $\omega_{0}=10\gamma$, and $\omega=J=\gamma$.}
	\label{Fig.2}
\end{figure}

First, we will assume that N two-level systems are located in a reservoir that follows fermionic statistics with the chemical potential $\mu$ and the temperature $T$, where the system particles collectively coupled to the environment and the system can exchange particles with the fermion reservoir. Note that in boson reservoirs (such as photon baths), particle number isn't conserved and chemical potential practically vanishes. Therefore, it is apparent that Bose-Einstein statistics are not applicable for examining the role of chemical potential in the charging process of open quantum batteries. The total system's Hamiltonian is expressed as $H=H_{S}+H_{R}+H_{int}$, with
\begin{align}
H_{R}=\sum_{k}\omega_{k}~c^{\dagger}_{k} c_{k},
\end{align}
and 
\begin{align}
H_{int}= \sum_{i=1}^{N}\sum_{k}f_{ik}(\sigma_{-}^{(i)}c^{\dagger}_{k}+\sigma_{+}^{(i)}c_{k})~.
\end{align}
In this context, $H_{R}$ represents the free Hamiltonian of the reservoir, where $c_{k}$ and $c^{\dagger}_{k}$ denote the annihilation and creation operators, respectively, for the $k$th mode with frequencies $\omega_{k}$ within the reservoir. $H_{int}$ signifies the qubit-reservoir interaction Hamiltonian while operating under the rotating wave approximation, and $f_{ik}$ denotes the real-valued qubit-reservoir coupling strengths for the $i$th qubits.

Quantum master equation for the reduced density operator of such a system under Born-Markov approximations in a weak-coupling regime is given by \cite{Damanet:1,bhattacharya:1}
\begin{align}\label{3}
\frac{d\rho}{dt}=D[\rho]= -i~[H_{S},\rho]+\mathcal{D_{-}}(\rho) + \mathcal{D_{+}}(\rho)~,
\end{align}
where 
\begin{align}\label{4}
	\mathcal{D_{-}}(\rho)=\sum_{i,j=1}^{N}\gamma~(1-n(\omega))~(\sigma_{j}^{-}\rho~\sigma_{i}^{+}-\frac{1}{2}\{\sigma_{i}^{+}\sigma_{j}^{-},\rho\})~,
\end{align}
and 
\begin{align}
\mathcal{D_{+}}(\rho)=\sum_{i,j=1}^{N}\gamma~n(\omega)~(\sigma_{j}^{+}\rho~\sigma_{i}^{-}-\frac{1}{2}\{\sigma_{i}^{-}\sigma_{j}^{+},\rho\})~,
\end{align}
describe the spontaneous and thermally induced emission (dissipation) and thermally induced absorption (incoherent driving) processes, respectively. In the above, $\gamma$ is frequency-independent spectral density of the reservoir in contact with the subsystems. $n(\omega)=[\exp{(\omega-\mu)/T}+1]^{-1}$ is the average particle number on frequency $\omega$ in the fermion reservoir, which follows the Fermi-Dirac statistics. Note that, the
system can exchange particles with the fermion reservoirs in processes conserving the particle number (e.g., in quantum dot systems). In the long-time limit, the system relaxes to the steady-state regardless of the initial conditions. One way to achieve a dynamic steady-state is to set the derivative of $\rho$ with respect to time to zero in Eq.~\eqref{3}, denoted as $\dot{\rho}=0$. Finding the solution to the algebraic equation $D[\rho]=0$ is equivalent to identifying the eigenvector of the linear map $D[\cdot]$ corresponding to a zero eigenvalue. The matrix representation of terms $\mathcal{M}\rho\mathcal{N}$ in the Liouvillian $D[\cdot]$ is represented by the Kronecker product $\mathcal{M}\otimes\mathcal{N}^{\dagger}\cdot\vec{\rho}$, where $\vec{\rho}$ is the column vector formed by appending the rows of the density matrix $\rho$ and then transposing the result. We utilize the QuTip library, which offers various built-in tools for solving master equations in Lindblad form \cite{JOHANSSON:1,JOHANSSON:2}. 

In the following, we investigate the ergotropy $\mathcal{W}$ in relation to particle exchange (chemical potential) for different values of temperature by Fig.~\ref{Fig.2}. It can be seen the ergotropy is a monotonically increasing function of the reservoir chemical potential, with a value of zero when $\mu\leq\omega$. In the other words, the chemical potential dependence reaches its minimum value at $\mu=\omega$ and exhibits a monotonic change as $\mu$ crosses $\omega$. Excitingly, the increase in reservoir temperature leads to a reduction in ergotropy due to the thermal effects. Therefore, in this equilibrium fermion reservoir setting, the full-charged quantum battery can be approached as $T\rightarrow 0$ at $\mu\gg\omega$ and fixed $J$. Ultimately, it is worth noting that the steady-state ergotropy behavior suggests reliable and stable charging of a quantum battery without any loss or deterioration due to the decay effects of the environment in a equilibrium setting.

\subsection{non-equilibrium case}
\begin{figure}[t] 
	\includegraphics[scale=0.45]{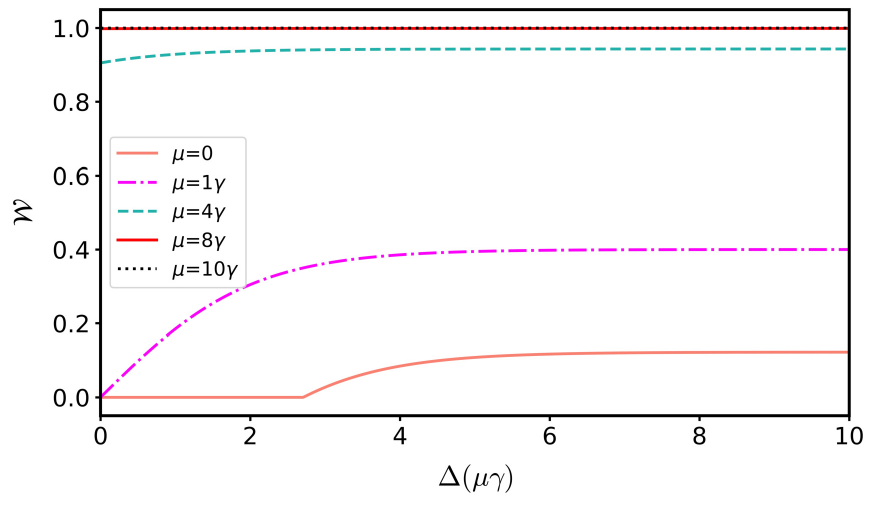}
	\caption{The ergotropy for the non-equilibrium case as a function of $\Delta(\mu\gamma)$ for different values of $\mu_{1}=\mu$ at $T=\gamma$. Other parameters are the same as in Fig.~\ref{Fig.2}.}
	\label{Fig.4}
\end{figure}

\begin{figure}[t] 
	\includegraphics[scale=0.32]{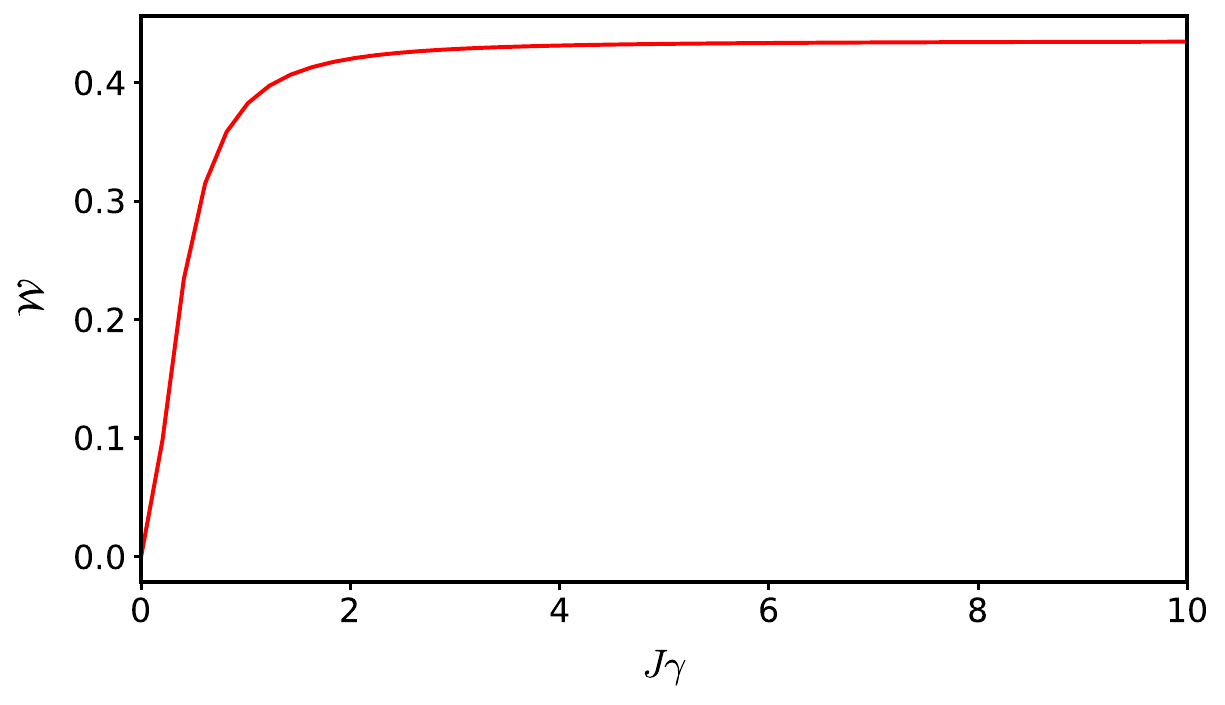}
	\caption{The ergotropy for the non-equilibrium case as a function of $J\gamma$ at $\mu=\gamma$, $\Delta\mu=2\gamma$ and $T=\gamma$. Other parameters are the same as in Fig.~\ref{Fig.2}.}
	\label{Fig.5}
\end{figure}

\begin{figure}[t] 
	\includegraphics[scale=0.32]{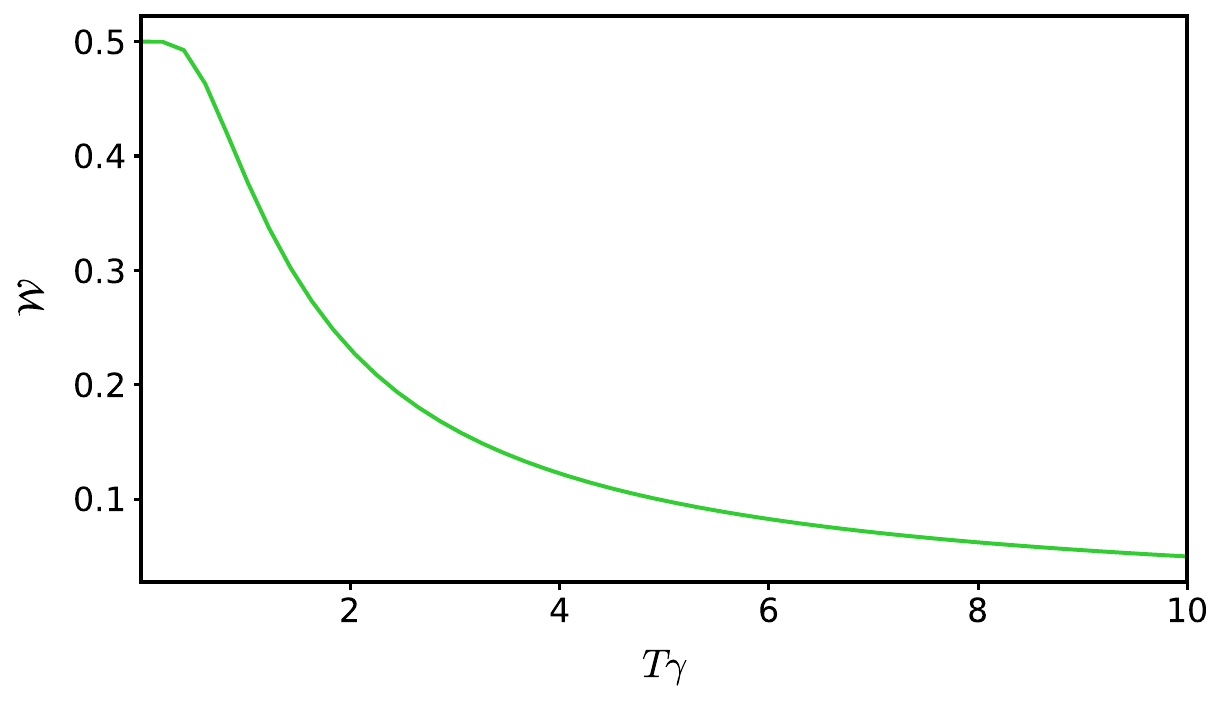}
	\caption{The ergotropy for the non-equilibrium case as a function of the temperature $T\gamma$ at $J=\gamma$. Other parameters are the same as in Fig.~\ref{Fig.5}.}
	\label{Fig.7}
\end{figure}

The non-equilibrium system under consideration composed of the N coupled
identical qubits that are immersed in their own fermion reservoirs with different chemical potentials and temperatures, where we are assuming individual decay for each qubit. The Hamiltonian for the total system reads $H=H_{S}+H_{R}+H_{int}$, with
\begin{align}
	H_{R}= \sum_{j=1}^{N}\sum_{k}\omega_{jk}~b^{\dagger}_{jk} b_{jk},
\end{align}
and 
\begin{align}
	H_{int}= \sum_{i=j}^{N}\sum_{k}g_{jk}(\sigma_{-}^{(j)}b^{\dagger}_{jk}+\sigma_{+}^{(j)}b_{jk})~.
\end{align}
$H_{R}$ is the free Hamiltonian of the reservoirs, where $b_{jk}$ and $b^{\dagger}_{jk}$
are the annihilation and creation operators, respectively, for the $k$th mode with frequencies $\omega_{jk}$ in the $j$th reservoir. $H_{int}$ is the qubit-reservoir interaction Hamiltonian under the rotating wave approximation, and $g_{jk}$ is the qubit-reservoir coupling strengths assumed to be real.

The dissipative dynamics of the system is described by a master equation of the form \eqref{3} with \cite{bhattacharya:1,Latune:1}
\begin{align}
	\mathcal{D_{-}}(\rho)=\sum_{i=1}^{N}\gamma~(1-n_{i}(\omega))~(\sigma_{i}^{-}\rho~\sigma_{i}^{+}-\frac{1}{2}\{\sigma_{i}^{+}\sigma_{i}^{-},\rho\})~,
\end{align}
and 
\begin{align}
	\mathcal{D_{+}}(\rho)=\sum_{i=1}^{N}\gamma~n_{i}(\omega)~(\sigma_{i}^{+}\rho~\sigma_{i}^{-}-\frac{1}{2}\{\sigma_{i}^{-}\sigma_{i}^{+},\rho\})~,
\end{align}
where $n_{i}(\omega)=[\exp{(\omega-\mu_{i})/T_{i}}+1]^{-1}$ is the average particle number on frequency $\omega$, with $\mu_{i}$ and $T_{i}$ being the chemical potential and the temperature of the $i$th reservoir, respectively.

\begin{figure}[t] 
	\includegraphics[scale=0.4]{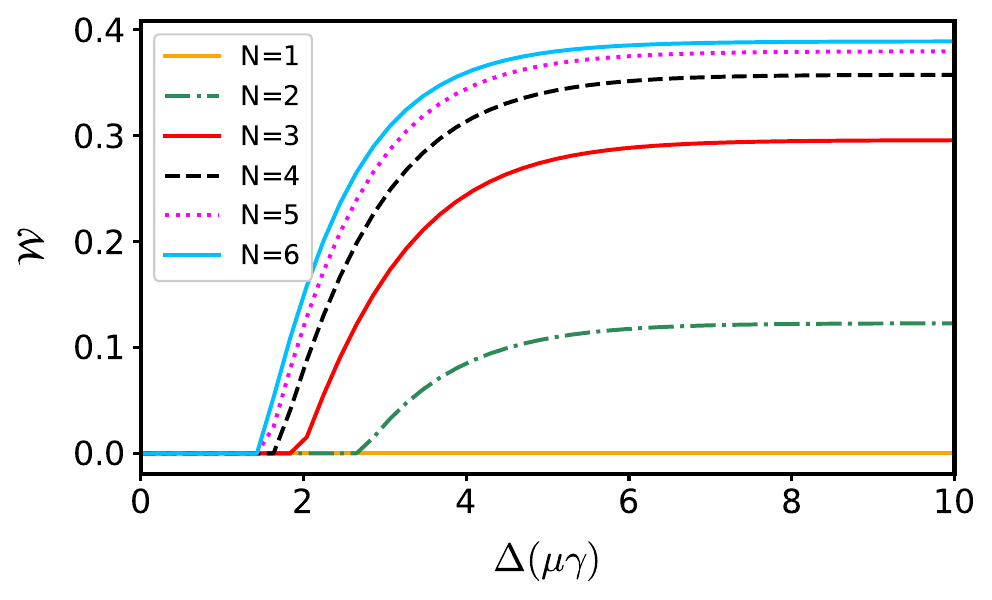}
	\caption{The ergotropy for the non-equilibrium case as a function of $\Delta(\mu\gamma)$ for different values of number cells $N$ at $T=\gamma$ and $\mu=0$. Other parameters are the same as in Fig.~\ref{Fig.2}.}
	\label{Fig.6}
\end{figure}

As a special case, we consider the N coupled qubits in contact with their individual fermion reservoirs with the same temperature $T_{i}=T$ but different chemical potentials $\mu_{1}=\mu$ and $\mu_{i}=\mu+\Delta\mu~(i=2,...,N)$. The non-equilibrium condition is characterized by the chemical
potential difference $\Delta\mu$. The ergotropy $\mathcal{W}$ as functions of $\Delta\mu$ for different values of $\mu_{1}=\mu$ is plotted in Fig.~\ref{Fig.4}. The parameters of $\omega$, $J$, and $\gamma$ are the same as those for the equilibrium baths case. As can be seen ergotropy has a manifest behavior with respect to $\Delta\mu$, depending on the value of the base chemical potential $\mu_{1}$. For $\mu_{1}\geq \omega$ the ergotropy monotonically increases from its equilibrium value as increases $\Delta\mu$. As $\mu_{1}$ becomes large enough (e.g., $\mu_{1}=8$ ), the ergotropy is significantly maximized no matter how large $\Delta\mu$ is. Also, the ergotropy also approaches some fixed values as $\Delta\mu$ grows large enough. The graph of $\mathcal{W}$ as function of the coupling strength $J$ is shown in Fig.~\ref{Fig.5}, with the temperature fixed at a relatively low value $T=1$. As can be expected the ergotropy increase with the inter-qubit coupling strength in general. Although, when $J$ is minor, the ergotropy is greatly reduced, however, for larger couplings (e.g $J\geq\omega$), regardless of their values, the ergotropy displays similar trends with respect to $J$. Our results also indicate that the thermal effect caused by the temperature of the reservoirs has a destructive effect on the maximum charge of the battery, so that at high temperatures limit the battery remains almost empty (see Fig.~\ref{Fig.7}).  Surprisingly, the difference between the role of the thermodynamic variables ($T$ and $\mu$) in Fermi-Dirac statistics has primarily contributed to this result. Additionally, in Fig.~\ref{Fig.6}, we observe a significant advantage in the number of auxiliary cells used for quantum battery charging. It is evident that for small values of N, even an increment of one cell will significantly shift the maximum ergotropy, making it impossible to charge the battery without the assistance of an ancilla. As the value of N decreases, a stronger bias is required to charge the battery. However, this sensitivity diminishes for larger values of cell number. 

Finally, judging from the maximum ergotropy that can be asymptotically achieved or approached, the equilibrium and non-equilibrium fermion reservoirs with particle exchange enhance quantum battery charging without requiring an external charging field. We note that the steady-state ergotropy demonstrates a straightforward pattern in the present scenarios, allowing the battery to be charged consistently and sustainably without initial restrictions, while also safeguarding it against environmental influences, such as self-discharging.

\section{Conclusion}\label{IV}
In this paper, using the Born-Markov master equation, we studied the steady-state charging process of a single-cell quantum battery embedded in the center of a star network comprising N-qubits, under two different equilibrium and non-equilibrium scenarios. We derived numerical solutions for the dynamic steady-state, enabling us to analyze ergotropy behaviors when taking the partial trace over surrounding qubits. In the equilibrium case, we found that the ergotropy with the chemical potential grows monotonically and for some parameter regimes it is independent of the coupling strength or the number of qubits. In the non-equilibrium case, ergotropy behaves as a monotonic function with non-equilibrium conditions (chemical potential difference) and the high base chemical potential of the reservoir corresponding to the quantum battery strongly boosts the steady-state charging regardless of the chemical potential difference. Whereas if the base temperature is high enough or the coupling strength between qubits is weak, ergotropy is strongly suppressed regardless of the strength of the non-equilibrium condition. Furthermore, our research shows that when qubits are coupled to fermion reservoirs that exchange particles with the system, there is a significant improvement in quantum battery charging.

Our results suggest some efficient strategies that may be useful for optimizing the steady-state charging of open quantum batteries and mitigating the detrimental impact of self-discharging without requiring auxiliary charging fields. Among these, one can address the use of fermion reservoirs with a low base temperature, strong enough coupling strength between qubits, establishing non-equilibrium conditions by chemical potential difference, connecting qubits with a higher transition frequency to reservoirs with a lower frequency, and connecting the battery to reservoirs with a relatively high base chemical potential in non-equilibrium situations. Therefore, present findings may catalyze future exploration of an innovative approach to environment-mediated charging, eliminating the need for an external charging field—a topic that has received limited research attention until now.
In the other words, the general guidelines outlined here have the potential to facilitate additional research on steady-state ergotropy in the field of quantum batteries, enabling evaluation across diverse scenarios.
%\newpage
\begin{acknowledgments} 
S. Salimi thanks research funded by Iran national science foundation (INSF) under project No.4003162.
\end{acknowledgments}
%%%%%%%%%%%%%%%%%%%%%%%%%%%%%
%\appendix
%%%%%%%%%%%%%%%%%%%%%%%%%%%%%

%%%%%%%%%%%%%%%%%%%%%%%%%%%%%

%\section{}\label{appA}

\label{apendice-prova-MSfases}
\bibliography{mybib-URL.bib}

%\bibliography{/home/acsantos/Dropbox/School/Articles/Gaveta/Models/Bibliografia/mybib-URL.bib}

\end{document}